\documentclass[aps,reprint,groupedaddress,showpacs]{revtex4-1}
\usepackage{amssymb}
\usepackage{amsmath}
\usepackage{dcolumn}
\usepackage{bm}
\usepackage{graphicx}
\usepackage{mathrsfs}

\begin{document}

\title{Robust continuous-variable entanglement of microwave photons  with cavity electromechanics}
\author{Peng-Bo Li}
\email{lipengbo@mail.xjtu.edu.cn}
\homepage{http://lipengbo.gr.xjtu.edu.cn/}
\author{Shao-Yan Gao}
\author{Fu-Li Li}
\affiliation {MOE Key Laboratory for Nonequilibrium Synthesis and Modulation of Condensed Matter,\\
Department of Applied Physics, Xi'an Jiaotong University, Xi'an
710049, China}

\begin{abstract}
We investigate the controllable generation of robust photon entanglement with a  circuit cavity electromechanical system, consisting of two superconducting coplanar waveguide cavities (CPWC's) capacitively coupled by a nanoscale mechanical resonator (MR).
We show that, with this electromechanical system, two-mode continuous-variable entanglement of cavity photons  can be
engineered deterministically either via coherent
control on the dynamics of the system, or through a dissipative quantum dynamical process.
The first scheme, operating in the strong coupling regime, explores the excitation of the cavity Bogoliubov modes, and is insensitive to the initial thermal noise. The
second one is based on the reservoir-engineering approach, which exploits the mechanical dissipation as a useful resource to
perform ground state cooling of two delocalized cavity Bogoliubov modes.
The achieved amount of entanglement in both schemes is determined by the relative ratio of the effective electromechanical coupling
strengths, which thus can be tuned and made much lager than that in previous studies.

\end{abstract}
\pacs{42.50.Pq, 85.85.+j, 03.67.Bg, 85.25.-j}
\maketitle

\section{Introduction}

Circuit cavity electromechanics \cite{JPCS-264-012025,nature-471-204,CRP-13,PR-511}, the counterpart of cavity optomechanics \cite{SCI-321-1172,Phys-2} in the form of electrical circuits, describes
the parametrical coupling between the motion of a micro or nanoscale MR and an electrical circuit.
The underlying physics of cavity electromechanics is that the motion of the mechanical oscillator modulates the capacitance
of the electrical circuit, thus creating parametrical coupling between these two systems. Compared to their optomechanical analogues, electromechanical systems have the advantages that these low-loss superconducting circuits are easily cooled to ultralow temperatures \cite{nature-464-697},
and can be fabricated on a single chip using the standard optical lithographic techniques.
Recent experimental and theoretical progress has shown that cavity optomechanics and electromechanics are pretty
useful for macroscopic tests of the fundamental laws of quantum mechanics, or for other practical applications relevant with quantum phenomena \cite{nature-480-351,natphys-6-602,prl-99-093901,prl-99-093902,prl-101-263602,prl-105-220501,prl-107-133601,prl-109-147205,prl-109-147206,natphys-4-555,OE-19-24905,Natcomm-3-987,prl-107-273601,prl-93-070501,prl-109-013603,prl-109-063601,prl-109-223601,prl-101-197203,NJP-10-095002,prb-84-054503,apl-101-141905,apl-101-063102,prb-86-155448,prl-108-033602}. Of particular interest is  the generation of non-classical motional, photonic and hybrid quantum states for basic tests of
quantum theory, as well as applications in quantum information processing \cite{prl-88-148301,prl-97-267201,prl-101-200503,NJP-10-115001,NJP-14-075014,NJP-14-125005,NJP-10-095010,pra-76-042336,pra-78-062303,prb-70-205304,pra-85-033805,pra-85-033822,prb-85-205415,epl-93-18003,pra-86-042306}.

In order to achieve entanglement of photons, one can use a  three-mode
optomechanical system, consisting of two optical target modes and a mechanical auxiliary mode.
Several theoretical works have described such schemes for entanglement generation, which use the auxiliary mode to mediate an
effective coherent interaction between the two target modes \cite{prl-99-250401,prl-109-130503,pra-86-013809,pra-79-024301,pra-84-042342}.
However, the entanglement generated in those protocols often suffers from the unavoidable decoherence and dissipation  associated with
such systems. For instance, the mechanical thermal noise and mechanical dissipation available in optomechanical system
often play a negative role in the entanglement preparation process. The traditional method for
beating such decoherence process often needs the strong  photon-phonon interaction to exceed the decay of the
photons and phonons. The achieved photon entanglement is often limited by the constraints on the magnitude of the
optomechanical coupling strengths. It is thus appealing to present some new schemes for robust photon entanglement with such three-mode
optomechanical or electromechanical systems.

In this work, we study the robust generation of photon entanglement with a circuit cavity electromechanical system consisting of two superconducting
CPWC's and a nanoscale or micro MR. In the proposed experimental setup, the superconducting cavities are capacitively
coupled by a capacitor that incorporates the nanoscale or micro MR into its electrode plates, and is biased by
a driving voltage. In this case, the cavity modes only couple with the mechanical mode and do not interact with each other. We show that, through suitably choosing the driving frequencies of the voltages, we can generate various linear operations
between cavity photons and mechanical phonons on demand via the modulation of the coupling
strength.

In particular, with this circuit electromechanical system we present two different protocols for
preparing continuous-variable entangled states of microwave photons.
The first protocol, operating in the strong coupling regime, relies on coherent
control over the dynamics of the system and is insensitive to the initial thermal noise.
In this scheme, we show that at selected time the Bogoliubov  modes  composed of the cavity modes only can be excited.
Since these cavity Bogoliubov modes do not contain the mechanical mode, the scheme is hence
robust against the mechanical noise. The second protocol is based on
a dissipative quantum dynamical process, which exploits the mechanical dissipation as a resource and only needs high frequency low-Q mechanical oscillators. We show that at steady state robust entanglement of two cavity modes can be generated through tailoring the
dissipative environment of the two target modes. In the present case, the reservoir-engineering scheme
exploits the ground state cooling of two delocalized cavity Bogoliubov modes.
The photon entanglement achieved in both schemes is determined by the ratio of the effective electromechanical coupling strengths, rather than their
magnitudes. Therefore, the amount of entanglement obtained in these schemes can be far greater than that in previous works.
These protocols may have promising applications for continuous-variable quantum information processing with cavity
electromechanics.

\section{Coupling  two superconducting CPWC's via a MR}
\begin{figure}[h]
\centerline{\includegraphics[bb=20 503 588 786,totalheight=1.5in,clip]{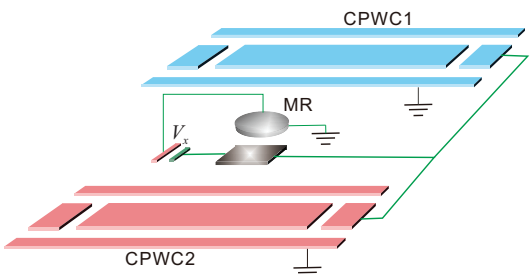}}
\caption{(Color online) The schematic of  two superconducting CPWC's capacitively coupled by a nanoscale MR, which is driven by a gate voltage $V_x$.}
\end{figure}
As shown in Fig.1, we consider an electromechanical system where two superconducting CPWC's are capacitively coupled by a nanoscale or micro MR, which is driven by a gate voltage $V_x$.
To implement this scheme, the superconducting CPWC's are electrically connected with a capacitor biased by a driving voltage.
The capacitor is formed by two parallel metal plates, one of which is replaced by a metallic membrane realizing the MR (drum resonator) \cite{nature-471-204}.
Alternatively, one can choose to use a micromechanical bulk dilatational resonator, as recently used in the experiment to couple with a phase qubit \cite{nature-464-697}. The working frequency of these MR's is in the range of GHz, and they can couple to the CPWC's through interdigitated capacitors.
In  both cases, the capacitance $C$ of the capacitor  is dependent on the MR displacement $X=\sqrt{\hbar/(2m\omega_m)}(\hat{b}+\hat{b}^\dag)$,
where $m$ is the mass of the MR, $\omega_m$ the mechanical vibration frequency, and $\hat{b}$  the annihilation operator for the MR.  If
we assume that the displacement is much smaller than the equilibrium distance
$d$ between the metallic membrane and the metallic base electrode, then the capacitance
approximately becomes $C=C_0(1+X/d)$, where $C_0$ is the capacitance for
the MR in equilibrium.

For a superconducting CPWC \cite{pra-69-062320}, the voltage at position $x$ is
\begin{eqnarray}
  V_j(x) &=& \sqrt{\frac{\hbar \omega_j}{C_j}}(\hat{a}_j^\dag+\hat{a}_j)\cos(2\pi x/L_j), (j=1,2),
\end{eqnarray}
where $\omega_j$ is the resonant frequency, $C_j$ the total capacitance,  $\hat{a}_j$ the annihilation operator, and $L_j$ the length for the $j$th CPWC respectively. With a coupling capacitance $C$ between these CPWC's, the coupled interaction can be derived as
\begin{eqnarray}
\label{H1}
  H_I&=&\frac{1}{2}C_0(1+X/d)(V_1(0)+V_2(0)-V_x)^2.
\end{eqnarray}
We subsequently perform a rotating-wave approximation to simplify the coupled interaction.
After neglecting rapidly oscillating and other higher order terms, the Hamiltonian describing the coupled system can be derived
as
\begin{eqnarray}
\label{H2}
\mathscr{H}&=&\hbar\omega_1\hat{a}_1^\dag\hat{a}_1+\hbar\omega_2\hat{a}_2^\dag\hat{a}_2+\hbar\omega_m\hat{b}^\dag\hat{b}\nonumber\\
&&-\hbar g_1(t)(\hat{b}+\hat{b}^\dag)(\hat{a}_1+\hat{a}_1^\dag)-
\hbar g_2(t)(\hat{b}+\hat{b}^\dag)(\hat{a}_2+\hat{a}_2^\dag)
\end{eqnarray}
where $g_j(t)=\frac{C_0}{d}\sqrt{\frac{\omega_j}{2m\omega_mC_j}}V_x(t)$, and we have included
the free Hamiltonian of the two CPWC's and MR in the first three terms. The last two terms in Eq. (\ref{H2}) are the linear interaction between the CPWC's and the MR. Up to now the result is valid for arbitrary driving voltage signals. Through adjusting the driving frequency of the voltage, we can generate various linear operations.
For instance, the beam-splitter interaction between two cavity modes and the mechanical mode,  as discussed in Ref. \cite{prl-108-153603,prl-108-153604} to realize intracavity state transfer, can be recovered from  Hamiltonian (\ref{H2}) by choosing $V_x(t)=V_x^0\cos \omega_dt$ and setting $\omega_d=\omega_1-\omega_m=\omega_2-\omega_m$.
In this way, the modulation of the coupling
strength provides an effective tool for controlling the interaction between the CPWC's and the MR.
In the following section, we will show how to prepare robust two-mode entangled states of the cavity photons through
engineering the desired interaction between photons and phonons.

\section{Generating continuous-variable entanglement of photons confined in the CPWC's}
We now consider the case where the MR is driven by a gate voltage of the form $V_x(t)=V_x^1\cos(\omega_d^1t)+V_x^2\cos(\omega_d^2t+\phi)$, where
$\phi$ is a fixed phase difference between the voltage components.
If we choose the driving frequencies as $\omega_d^1=\omega_1+\omega_m,\omega_d^2=\vert\omega_m-\omega_2\vert$, corresponding to the
blue sideband and red sideband driving for the MR, then under the rotating-wave approximation we can obtain the Hamiltonian in the
interaction picture
\begin{eqnarray}
\label{H3}
\mathscr{H}&=-\hbar \Theta_1 (\hat{a}_1^\dag \hat{b}^\dag+\hat{a}_1 \hat{b})-\hbar \Theta_2 (\hat{a}_2^\dag \hat{b}+\hat{a}_2 \hat{b}^\dag),
\end{eqnarray}
where
\begin{equation*}
    \Theta_j=\frac{C_0}{2d}\sqrt{\frac{\omega_j}{2m\omega_mC_j}}V_x^j,\quad(j=1,2).
\end{equation*}
The Hamiltonian (\ref{H3}) describes a
system of three coupled harmonic oscillators with controllable
coefficients \cite{pra-81-035802,pra-68-062317,JOP}. The first term describes simultaneous creation
or annihilation of a photon in CPWC1 and a phonon and is responsible for
entangling the CPWC1 and the MR, while the second
term describes  the exchange of excitation
quanta between the CPWC2 and the motion. These terms together will lead the CPWC's
to be entangled with each other. In what follows, we will discuss two different
schemes to realize this goal, one of which is based on coherent
control on the dynamics of the system, while the other is  through a dissipative quantum dynamical process.

\subsection{Dynamical generation of photon entanglement via excitations of the cavity Bogoliubov  modes}
We first focus on the regime where the dissipative effects on the coherent dynamics can be
neglected, i.e., the strong coupling regime, $\{\Theta_1,\Theta_2\}\gg \{\kappa_1,\kappa_2,n_{th} \gamma_m,\gamma_m\}$, where
$\kappa_j$ is the $j$th CPWC field decay rate, $n_{th}$ is the thermal equilibrium occupation
number for the mechanical mode at temperature $T$, and $\gamma_m$ is mechanical
dissipation rate. This regime can be easily realized, since the coupling strength $\Theta_j$ can be
tuned by the classical driving amplitude $V_x^j$, and high-Q superconducting CPWC's and MR's can be conveniently fabricated in the laboratory. In this limit the coherent dynamics of the coupled system can be
easily solved in the Heisenberg representation.

The Heisenberg equations of motion read
\begin{eqnarray}
  \dot{\hat{a}}_1 &=& i\Theta_1 \hat{b}^\dag, \\
  \dot{\hat{a}}_2 &=& i\Theta_2 \hat{b},\\
  \dot{\hat{b}} &=& i\Theta_1 \hat{a}_1^\dag+i\Theta_2 \hat{a}_2,
\end{eqnarray}
which would generate periodic dynamics provided that $|\Theta_2|>|\Theta_1|$.
After some straightforward derivations, we can obtain the time evolution of
the operators as
\begin{eqnarray}
\label {S1}
\hat{a}_1(t)&=&\frac{i\Theta_1}{\Theta}\hat{b}^\dag(0)\sin\Theta
t+\frac{\Theta_1\Theta_2}{\Theta^2}[1-\cos\Theta
t]\hat{a}_2^\dag(0)+\frac{1}{\Theta^2}[|\Theta_2|^2\nonumber\\
&&-|\Theta_1|^2\cos\Theta
t]\hat{a}_1(0),\\
\label {S2}
\hat{a}_2(t)&=&\frac{i\Theta_2}{\Theta}\hat{b}(0)\sin\Theta
t-\frac{\Theta_1\Theta_2}{\Theta^2}[1-\cos\Theta
t]\hat{a}_1^\dag(0)-\frac{1}{\Theta^2}[|\Theta_1|^2\nonumber\\
&&-|\Theta_2|^2\cos\Theta
t]\hat{a}_2(0),\\
\label {S3}
\hat{b}(t)&=&\hat{b}(0)\cos\Theta
t+\frac{1}{\Theta}[i\Theta_2\hat{a}_2(0)+i\Theta_1\hat{a}_1^\dag(0)]\sin\Theta
t,
\end{eqnarray}
with $\Theta=\sqrt{|\Theta_2|^2-|\Theta_1|^2}$. In general these solutions describe tripartite entanglement
among cavity modes and the mechanical oscillator. However, we find at the instant $T_{\pi}=\pi/\Theta$
Eqs. (\ref{S1})-(\ref{S3}) become
\begin{eqnarray}
\label {S4}
\hat{a}_1(T_{\pi})&=&\frac{|\Theta_1|^2
+|\Theta_2|^2}{\Theta^2}\hat{a}_1(0)+\frac{2\Theta_1\Theta_2}{\Theta^2}\hat{a}_2^\dag(0),\\
\label {S5}
\hat{a}_2(T_{\pi})&=&-\frac{|\Theta_1|^2
+|\Theta_2|^2}{\Theta^2}\hat{a}_2(0)-\frac{2\Theta_1\Theta_2}{\Theta^2}\hat{a}_1^\dag(0),\\
\label {S6}
\hat{b}(T_{\pi})&=&-\hat{b}(0).
\end{eqnarray}
Therefore, at this instant the mechanical motion is decoupled from the cavity modes and returns to its initial state.
Moveover, at the time $T_\pi$ the two cavity modes are entangled with each other.

To be more specific,
we introduce the unitary operator $\hat{S}(\zeta)=e^{\zeta\hat{a}_1\hat{a}_2-\zeta\hat{a}_1^\dag\hat{a}_2^\dag}$,
where $\zeta=\tanh^{-1} [2r/(1+r^2)],r=|\Theta_2/\Theta_1|$. Then we find Eqs. (\ref{S4}) and (\ref{S5}) can be
rewritten as the delocalized cavity Bogoliubov
mode operators
\begin{eqnarray}
\label {S7}
\hat{a}_1(T_{\pi})&=&\cosh \zeta\hat{a}_1(0)+\sinh \zeta\hat{a}_2^\dag(0)=\hat{S}\hat{a}_1(0)\hat{S}^\dag,\\
\label {S8}
\hat{a}_2(T_{\pi})&=&-(\cosh \zeta\hat{a}_2(0)+\sinh \zeta\hat{a}_1^\dag(0))=e^{i\pi}\hat{S}\hat{a}_2(0)\hat{S}^\dag.
\end{eqnarray}
These results imply that at the instant $T_\pi$ the  Bogoliubov  modes  composed of the cavity modes only will be excited.
Since these cavity Bogoliubov modes do not contain the mechanical mode, the scheme is hence
robust against the mechanical noise.

In the Schr\"{o}dinger picture, the time evolution operator of the total system corresponds to
$\hat{U}(T_\pi)=\hat{S}(\zeta)^\dag\otimes \hat{I}_m$, where $\hat{I}_m$ is the identity operator for the mechanical mode.
Thus, if initially the MR density matrix is a thermal state at
temperature $T$ given by
\begin{equation}
    \varrho_m(0)=(1-e^{-\hbar\omega_m/k_BT})e^{-H_m/k_BT}
\end{equation}
where $k_B$ is the Boltzmann constant, and $H_m=\hbar\omega_m(\hat{b}^\dag\hat{b}+1/2)$, then
$\varrho_m(T_\pi)=\varrho_m(0)$ in the Schr\"{o}dinger representation. Moveover, at the time
$T_\pi$, the  two cavity modes will be prepared in a two-mode squeezed state if the initial state  is the vacuum state for both
cavity modes, $\vert 00\rangle_c$. In particular, using the factored form of the two-mode
squeeze operator \cite{pra-31-3093}
\begin{eqnarray}
   \hat{S}(\zeta)&=&(\cosh \zeta)^{-1}e^{-\hat{a}_1^\dag\hat{a}_2^\dag\tanh \zeta}e^{-(\hat{a}_1^\dag\hat{a}_1+\hat{a}_2^\dag\hat{a}_2)\ln(\cosh\zeta)}e^{\hat{a}_1\hat{a}_2\tanh \zeta},
\end{eqnarray}
at the time $T_\pi$ the state of the two cavity modes is
\begin{eqnarray}
\label{SQ}
  \vert \psi\rangle_c &=& \frac{1}{\cosh\zeta}\sum^{\infty}_{n=0}(\tanh \zeta)^n\vert n,n\rangle_c \nonumber \\
  &=&\left(\frac{1-r^2}{1+r^2}\right)\sum^{\infty}_{n=0}\left(\frac{2r}{1+r^2} \right)^n \vert n,n\rangle_c.
\end{eqnarray}
The state (\ref{SQ}) is a two-mode
squeezed state of the photon fields in the two cavities, which
exhibits Einstein-Podolsky-Rosen (EPR) entanglement. The mechanical motion of the MR
plays a fundamental role in establishing the entanglement, nevertheless the initial motional state does not affect
the efficiency of the scheme. The
degree of squeezing (squeeze parameter $\zeta$) and amount of entanglement  are determined by
the ratio of $\Theta_2$ to $\Theta_1$, which can be controlled on demand
through tuning the driving signals. When the squeezed state is generated at the time of $T_\pi$, we
switch off the couplings between the CPWC's and the MR. Then
the squeezed state can be preserved until the cavity fields are
coupled out.

\begin{figure}[h]
\centerline{\includegraphics[bb=5 10 588 520,totalheight=2.5in,clip]{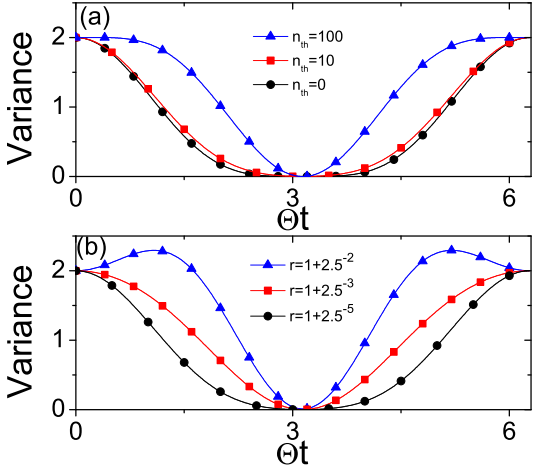}}
\caption{(Color online) The total
 variance $V$ versus $\Theta t$ for different values of the parameters $n_{th}$ and $r$: (a) $n_{th}=0,10,10^2$, $r=1+2.5^{-5}$; (b) $r=1+2.5^{-2},1+2.5^{-3},1+2.5^{-5}$, $n_{th}$=$10$.}
\end{figure}
In order to check the above analysis, we exploit the total variance $V=\langle(\Delta \hat{u})^2+(\Delta \hat{v})^2\rangle$ of a pair of EPR-like operators $\hat{u}=\hat{X}_1-\hat{X}_2$, and  $\hat{v}=\hat{P}_1+\hat{P}_2$ \cite{prl-84-2722}, with $\hat{X}_j=(\hat{a}_j+\hat{a}^\dag_j)/\sqrt{2}$, and $\hat{P}_j=-i(\hat{a}_j-\hat{a}^\dag_j)/\sqrt{2},j=1,2$. According to Ref. \cite{prl-84-2722}, a two-mode
Gaussian state is entangled if and only if $V<2$. For the two-mode  squeezed vacuum state $\hat{S}^\dag(\zeta)\vert00\rangle_c$, the total variance $V=\langle(\Delta \hat{u})^2+(\Delta \hat{v})^2\rangle=2e^{-2\zeta}$, implying this state exhibits EPR entanglement. In Fig. 2, the quantity $V$ is plotted
versus the scaled time $\Theta t$, for different values of the
initial thermal phonon number $n_{th}$ and the parameter $r$.
Fig. 2(a) shows the total variance $V$ as a function of the scaled time $\Theta t$ under different values of the  thermal phonon number $n_{th}$ for a fixed parameter $r$. From Fig. 2(a) it can be found that the total variance $V$ is unaffected by thermal noise at the instant $T_\pi$, i.e., at half period it is independent on the initial thermal phonon number $n_{th}$. This result can be understood since our discussions hold under the conditions
that the couplings of the system to the thermal reservoirs can be neglected. In such a case the coherent dynamics is purely governed by the
Hamiltonian (\ref{H3}) and  thermal effects enter only as an initial condition for the
mechanical mode. Fig. 2(b) displays $V$ versus $\Theta t$ under different values of the ratio $r$ with a fixed mean phonon number $n_{th}$.
We find that at half period the two CPWC's are steered into a two-mode squeezed state starting from a thermal state for the MR. The degree of
squeezing and the total variance for the two-mode squeezed state are determined by the parameter $r$. Therefore, the amount of entanglement can be tuned
and made large by the  ratio of the effective electromechanical coupling strengths.
These numerical results are in accordance with the analytical conclusions above.

\subsection{Robust photon entanglement at steady state through a dissipative dynamical process}
In the previous section, we have discussed how to prepare photon entanglement via coherent control on the evolution of the hybrid system. Though the protocol seems promising, the experimental implementation of this scheme requires stringent conditions, i.e., the coupling to the environment reservoirs should be neglected, thus requiring very high quality factors for both the CPWC's and MR's. For superconducting  CPWC's, this condition can be fulfilled since high-Q superconducting stripline
cavities are easy to be fabricated in the laboratory. However, as for the mechanical oscillator, in particular which has to be
incorporated into a capacitor in order to couple to the CPWC's, this requirement is too demanding. In addition, high frequency MR's are
required when the quantum regime is entered. However at present these GHz mechanical oscillators are plagued by very low quality factors.
It is known that MR performance degrades
considerably as the oscillating  frequency increases.
In this section, we will
present an alternative scheme which exploits the mechanical dissipation as a useful resource and only needs high frequency low-Q mechanical oscillators.

In what follows, the system-environment interaction is
assumed Markovian, and then is described by a master equation
in Lindblad form. We assume that the CPWC's couple with the vacuum bath, but the MR couples with a thermal bath.
Then the time evolution of the density
operator $\hat{\rho}$ for the whole system is described by the master
equation
\begin{eqnarray}
\label{M1}
\frac{d\hat{\rho}}{dt}&=&-\frac{i}{\hbar}[\mathscr{H},\hat{\rho}]+\mathscr{L}_{c_1}\hat{\rho}+\mathscr{L}_{c_2}\hat{\rho}+\mathscr{L}_m\hat{\rho},
\end{eqnarray}
where
\begin{eqnarray}
\mathscr{L}_{c_j}\hat{\rho}&=&\frac{\kappa_j}{2}(2\hat{a}_j\hat{\rho}\hat{a}^\dag_j-\hat{a}^\dag_j\hat{a}_j\hat{\rho}-\hat{\rho}\hat{a}^\dag_j\hat{a}_j),\nonumber\\
\mathscr{L}_m\hat{\rho}&=&\frac{\gamma_m}{2}(n_{th}+1)(2\hat{b}\hat{\rho}\hat{b}^\dag-\hat{b}^\dag\hat{b}\hat{\rho}-\hat{\rho}\hat{b}^\dag\hat{b})\nonumber\\&&+
\frac{\gamma_m}{2}n_{th}(2\hat{b}^\dag\hat{\rho}\hat{b}-\hat{b}\hat{b}^\dag\hat{\rho}-\hat{\rho}\hat{b}\hat{b}^\dag).
\end{eqnarray}
In the following we focus on the regime where $ \gamma_m\gg\{\Theta_1,\Theta_2\}\gg \{\kappa_1,\kappa_2,n_{th} \gamma_m\}$.
The condition $\gamma_m\gg\{\Theta_1,\Theta_2\}$ corresponds to strong mechanical damping for the MR, i.e., very low quality factors, while
$n_{th}\gamma_m\ll\gamma_m$ implies near zero temperature for the mechanical mode, thus requiring ground state cooling of the MR. In effect,
in the regime of large mechanical frequency (in the GHz range) and at cryogenic temperature, the thermal phonon number is nearly zero, i.e.,
$n_{th}=(e^{\hbar\omega_m/k_BT}-1)^{-1}\simeq0$, which corresponds to coupling with the vacuum bath for the MR. Under this regime the master equation (\ref{M1})
then can be approximated as
\begin{eqnarray}
\label{M2}
\frac{d\hat{\rho}}{dt}&=&-\frac{i}{\hbar}[\mathscr{H},\hat{\rho}]
+\frac{\gamma_m}{2}(2\hat{b}\hat{\rho}\hat{b}^\dag-\hat{b}^\dag\hat{b}\hat{\rho}-\hat{\rho}\hat{b}^\dag\hat{b}).
\end{eqnarray}

We now introduce the phonon number representation for the density operator $\hat{\rho}$, i.e., $\hat{\rho}=\sum_{m,n=0}^\infty\rho_{mn}\vert m\rangle\langle n\vert$, where $\rho_{mn}$ are the density-matrix elements in the basis of the phonon number states $\{\vert n\rangle,n=0,1,2,...\}$, and are still operators with respect to the cavity fields. Under the condition of strong mechanical damping, the populations of the highly excited motional states can be neglected. Therefore, we consider only the matrix elements $\rho_{mn}$ inside the subspace $\{\vert 0\rangle, \vert 1\rangle\}$ of the phonon numbers. In this case, the master equation (\ref{M2}) leads to the following set of coupled equations of motion for the density-matrix elements
\begin{eqnarray}
\label{M3}
\dot{\rho}_{00}&=&-i\rho_{01}(\Theta_1\hat{a}_1^\dag+\Theta_2\hat{a}_2)+i(\Theta_1\hat{a}_1+\Theta_2\hat{a}_2^\dag)\rho_{10}+\gamma_m\rho_{11},\\
\label{M4}\dot{\rho}_{11}&=&-i\rho_{10}(\Theta_1\hat{a}_1+\Theta_2\hat{a}_2^\dag)+i(\Theta_1\hat{a}_1^\dag+\Theta_2\hat{a}_2)\rho_{01}-\gamma_m\rho_{11},\\
\label{M5}\dot{\rho}_{01}&=&-i\rho_{00}(\Theta_1\hat{a}_1+\Theta_2\hat{a}_2^\dag)+i(\Theta_1\hat{a}_1+\Theta_2\hat{a}_2^\dag)\rho_{11}-\frac{\gamma_m}{2}\rho_{01}.
\end{eqnarray}
In the regime of strong damping rate $\gamma_m$, the elements $\rho_{01}$ and $\rho_{11}$ can be adiabatically eliminated from the above equations, leading to
\begin{eqnarray}
\label{M6}\rho_{01}&=&\frac{2i\Theta}{\gamma_m}(\mathscr{D}^\dag\rho_{11}-\rho_{00}\mathscr{D}^\dag)
\end{eqnarray}
where
\begin{eqnarray}
\mathscr{D}&=&\frac{\Theta_2}{\Theta}\hat{a}_2+\frac{\Theta_1}{\Theta}\hat{a}_1^\dag.
\end{eqnarray}
is the cavity Bogoliubov mode operator.
The reduced density operator for the CPWC's can be approximated as  $\hat{\varrho}_c=\mbox{Tr}_m(\hat{\rho})\simeq\rho_{00}+\rho_{11}$.
Replacing (\ref{M6}) into (\ref{M3}) and (\ref{M4}), and adding up them, after neglecting higher-order terms we obtain
the evolution of the cavity modes with an effective master equation
\begin{eqnarray}
\label{M7}
\frac{d\hat{\varrho}_c}{dt}&=&\frac{\Gamma_c}{2}(2\mathscr{D}\hat{\varrho}_c\mathscr{D}^\dag-\mathscr{D}^\dag\mathscr{D}\hat{\varrho}_c
-\hat{\varrho}_c\mathscr{D}^\dag\mathscr{D})
\end{eqnarray}
with $\Gamma_c=4\Theta^2/\gamma_m$.
This master equation has the form of the standard engineering reservoir scheme, which describes ground state
cooling of the cavity  Bogoliubov mode $\mathscr{D}$. The only pure steady state of the
system is the eigenstate $\vert\psi\rangle$ of the operator $\mathscr{D}$ with zero eigenvalue, ensuring that there is
no further eigenstate $\vert \phi\rangle$ of $\mathscr{D}$ such that $[\mathscr{D},\mathscr{D}^\dag]\vert \phi\rangle=0$.
For the operator $\mathscr{D}=\frac{\Theta_2}{\Theta}\hat{a}_2+\frac{\Theta_1}{\Theta}\hat{a}_1^\dag$, we will
find that this condition cannot be satisfied. From the eigenvalue equation $\mathscr{D}\vert\psi\rangle=0 $, and the
relation $\mathscr{D}=\mathcal {S}\hat{a}_2\mathcal {S}^\dag$, with $\mathcal {S}(\varsigma)=e^{\varsigma\hat{a}_1\hat{a}_2-\varsigma\hat{a}_1^\dag\hat{a}_2^\dag}$, $\varsigma=\tanh^{-1}[\Theta_1/\Theta_2]$, one can readily find that
\begin{eqnarray}
  \vert\psi\rangle_c &=&\mathcal {S}\vert \mu,0\rangle_c
\end{eqnarray}
is a steady state of the master equation (\ref{M7}), but not the only one.
Here $\vert \mu,0\rangle_c$ denotes  an arbitrary state for the first CPWC mode and the vacuum state for the second one.

In order to steer the system into the two-mode squeezed state $\mathcal {S}\vert 0,0\rangle_c $, we need another
dissipative process together with the described one, leading to the effective master equation
\begin{eqnarray}
\label{M8}
\frac{d\hat{\varrho}_c}{dt}&=&\frac{\Gamma_c}{2}(2\tilde{\mathscr{D}}\hat{\varrho}_c\tilde{\mathscr{D}}^\dag-\tilde{\mathscr{D}}^\dag\tilde{\mathscr{D}}\hat{\varrho}_c
-\hat{\varrho}_c\tilde{\mathscr{D}}^\dag\tilde{\mathscr{D}})\nonumber\\
&&+\frac{\Gamma_c}{2}(2\mathscr{D}\hat{\varrho}_c\mathscr{D}^\dag-\mathscr{D}^\dag\mathscr{D}\hat{\varrho}_c
-\hat{\varrho}_c\mathscr{D}^\dag\mathscr{D})
\end{eqnarray}
where $\tilde{\mathscr{D}}=\mathcal {S}\hat{a}_1\mathcal {S}^\dag$ is the other delocalized cavity Bogoliubov mode operator . This master equation
describes simultaneous ground state cooling of the system in the transformed picture with the basis $\tilde{\mathscr{D}},\mathscr{D}$ \cite{prl-108-043602,pra-86-012318}.
In fact, one can find that $\tilde{\mathscr{D}}\mathcal {S}\vert 0,0\rangle_c=\mathscr{D}\mathcal {S}\vert 0,0\rangle_c=0$, and
$[\tilde{\mathscr{D}},\mathscr{D}]=0$. Therefore, the unique steady state of master equation (\ref{M8}) is just
$\mathcal {S}\vert 0,0\rangle_c$. This two-mode squeezed state of the photon fields confined in the two cavities exhibits EPR entanglement.
The amount of entanglement is solely determined by the ratio of the effective electromechanical coupling strengths, thus can be tuned and made much larger.

The Lindblad term with respect to the $\tilde{\mathscr{D}}$ operator can
be engineered from the  Hamiltonian
\begin{eqnarray}
\label{H4}
\mathscr{H}'&=-\hbar \Theta_1 (\hat{a}_1^\dag \hat{b}+\hat{a}_1 \hat{b}^\dag)-\hbar \Theta_2 (\hat{a}_2^\dag \hat{b}^\dag+\hat{a}_2 \hat{b}),
\end{eqnarray}
following the same reasoning as that for the $\mathscr{D}$ operator, with the driving frequencies chosen as $\omega_d^1=\vert\omega_m-\omega_1\vert$, and  $\omega_d^2=\omega_2+\omega_m$. However, for the case of
just one MR, we cannot get the master equation (\ref{M8}) to realize simultaneous cooling of both modes $\tilde{\mathscr{D}}$ and $\mathscr{D}$, since
both $\mathscr{H}$ and $\mathscr{H}'$ cannot be possessed simultaneously only through adjusting the driving frequencies for one
MR. In order to have both cooling processes, in $\tilde{\mathscr{D}}$ and $\mathscr{D}$, we can employ a stroboscopic cooling scheme.
In this approach, the system evolves during a time $t$ in $N$ cycles of duration
$\delta t=t/N$, while the driving parameters alternate between the
ones with respect to $\tilde{\mathscr{D}}$ and those of $ \mathscr{D}$.
The stroboscopic limit is valid provided that the time interval $\delta t$ is much smaller than $1/\Gamma$,
in which case the effective dynamics of the system is just as that described by the master equation (\ref{M8}).
Alternatively, one can couple the CPWC's with two MR's, each of which is driven by a bichromatic microwave signal to
induce sidebands in the CPWC-MR coupling. In this case, one can realize the Hamiltonian $\mathscr{H}$ for one MR, and simultaneously
have the Hamiltonian $\mathscr{H}'$ for the other. In the regime of strong mechanical damping for both MR's, one can
exploit the engineering reservoir scheme to get the effective master equation (\ref{M8}).

It is necessary to verify the model through numerical
simulations. To provide an example, here we consider the two-MR case, where
the dynamics of the system can be simulated by the following master equation
\begin{eqnarray}
\label{M9}
\frac{d\hat{\rho}}{dt}&=&-\frac{i}{\hbar}[\mathscr{H},\hat{\rho}]+\mathscr{L}_{c_1}\hat{\rho}+\mathscr{L}_{c_2}\hat{\rho}
+\mathscr{L}_{m_1}\hat{\rho}+\mathscr{L}_{m_2}\hat{\rho},
\end{eqnarray}
with
\begin{eqnarray}
\mathscr{H}&=&-\hbar \Theta_1 (\hat{a}_1^\dag \hat{b}^\dag_1+\hat{a}_1 \hat{b}_1)-\hbar \Theta_2 (\hat{a}_2^\dag \hat{b}_1+\hat{a}_2 \hat{b}^\dag_1)\nonumber\\
&&-\hbar \Theta_1 (\hat{a}_1^\dag \hat{b}_2+\hat{a}_1 \hat{b}^\dag_2)-\hbar \Theta_2 (\hat{a}_2^\dag \hat{b}^\dag_2+\hat{a}_2 \hat{b}_2)\\
\mathscr{L}_{m_j}\hat{\rho}&=&\frac{\gamma_{m_j}}{2}(n_{th}+1)(2\hat{b}_j\hat{\rho}\hat{b}_j^\dag-\hat{b}_j^\dag\hat{b}_j\hat{\rho}-\hat{\rho}\hat{b}_j^\dag\hat{b}_j)\nonumber\\&&+
\frac{\gamma_{m_j}}{2}n_{th}(2\hat{b}_j^\dag\hat{\rho}\hat{b}_j-\hat{b}_j\hat{b}_j^\dag\hat{\rho}-\hat{\rho}\hat{b}_j\hat{b}_j^\dag).
\end{eqnarray}

\begin{figure}[h]
\centerline{\includegraphics[bb=20 371 582 722,totalheight=2in,clip]{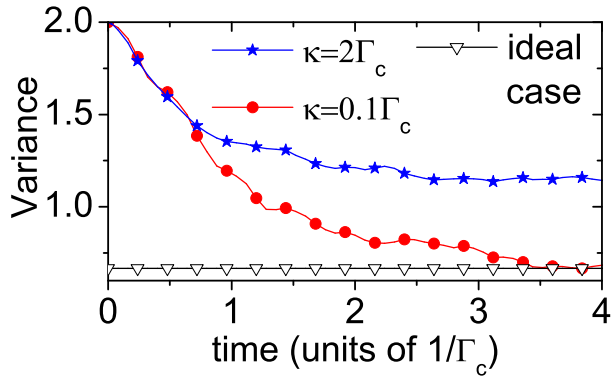}}
\caption{(Color online) Plot of time evolution of the total
variance $V$ through numerically solving the master equation (\ref{M9}), together with
the result for an ideal two-mode squeezed vacuum state. The relevant
parameters are chosen as $\Theta_2=2\Theta_1$, $\gamma_{m_1}\simeq\gamma_{m_2}\simeq\gamma_m=15\Theta_1,n_{th}=0.01$.}
\end{figure}
In Fig. 3 we illustrate the time evolution of the total
variance $V$ together with the result for an ideal two-mode  squeezed vacuum state,
under different values for the decay rate $\kappa$, where $\kappa_1\simeq\kappa_2\simeq\kappa$ is assumed. The initial state of the system is chosen as the ground states for the cavity modes and the mechanical mode.
The related parameters are chosen in such a way that they are within the parameter range for
which this scheme is valid and are accessible with current experimental setups.
From this figure we find that, when the effective decay rate for the engineered reservoir is much larger than the decay rate of the cavity photons, i.e., $\Gamma_c\gg\kappa$, at steady state nearly ideal EPR entanglement
($\Theta_2/\Theta_1=2,V=0.667$) between the photon fields in the
two CPWC's can be established with a fidelity of $99.98\%$. The time for reaching the stationary
state is about $T\simeq 4/\Gamma_c$. However, when $\kappa$ is of the same order of magnitude of or larger than $\Gamma_c$,
the ideal EPR entanglement is severely spoiled, since in this case the influence of the natural reservoir on the generated
two-mode squeezed state is almost the same
as that of the engineered reservoir. Thus, to reduce the
influence of the natural reservoir requires $\Gamma_c\gg\kappa$, or $4\Theta^2\gg\gamma_m\kappa$.
To make sure the protocol is valid, we also require that $\gamma_m\gg\Theta$. These conditions together imply that
$\gamma_m\gg\Theta\gg\sqrt{\gamma_m\kappa}/2$.

\section{Implementation}
Regarding the experimental feasibility of the proposals, currently available experimental setups of cavity
electromechanics \cite{nature-471-204,nature-480-351} are promising platforms for realizing the schemes.
We consider superconducting CPWC's with the fundamental frequency of $2\pi\times10$ GHz, whose damping rate can be as low as
$\kappa/2\pi\simeq10$ kHz given a quality factor $Q=10^6$ from recent circuit QED experiments.
The vacuum-fluctuations-induced voltage between the central conductor and the ground plane of the CPWC's is
typically of order of $\mu$V.
As for the nanoscale MR's,
we can choose to utilize an aluminium membrane integrated into a capacitor \cite{nature-471-204} for the first scheme, or a piezoelectric dilatation
resonator \cite{nature-464-697} for the second one, which comprises a piezoelectric thin film of aluminium nitride, sandwiched between two aluminium metal electrodes. The superconducting CPWC's and nano MR's can be
fabricated on a single chip with wafer-scale optical lithographic techniques.
For a nearly circular membrane with a diameter of 15 $\mu$m and a thickness of $100$ nm \cite{nature-471-204}, drum-like modes are allowed to resonate freely.
The fundamental mode is $\omega_m/2\pi=10.69$ MHz, giving a zero-point
motion of $4.1$ fm and a damping rate $\gamma_m/2\pi=30$ Hz \cite{nature-471-204}. For the first scheme, with the chosen parameters $V_x^1=10$ V, $V_x^2=1.01V_x^1$, $d=50$ nm, $C_0=40$ fF, we get $\Theta\sim3$ MHz, and the operation time for generating the target state is about $T_\pi\sim 1$ $\mu$s.
This time is much shorter than the photon life time, and the decoherence time for the mechanical mode with about $10^3$ phonons.
With regard to the second scheme, the piezoelectric dilatation
resonator is particularly suitable \cite{nature-464-697}, which has 6 GHz frequency and very strong damping rate $\gamma_m/2\pi\simeq23$ MHz ($Q\simeq260$). At
the temperature $T\simeq25$ mK, the number of  thermal phonons in the mechanical mode is less than $0.07$.
If we assume that $V_x^1=1$ V, $V_x^2=2V_x^1$, $d=50$ nm, $C_0=25$ fF, then we have $\Theta\sim17$ MHz.
The time for reaching the stationary state is about $4/\Gamma_c\sim0.5$ $\mu$s.

Finally, We discuss how to measure the entanglement between the
resonators. To implement this task one can use the experimental state tomography technique realized recently to detect a two-mode squeezed state in the microwave domain \cite{prl-107-113601}. In the experiment, all four quadrature components $X_1,X_2,P_1,P_2$ of a two-mode squeezed state are measured in a two-channel
heterodyne setup using amplitude detectors. Then, the full covariance matrix can be determined via analyzing two-dimensional phase space histograms for all possible pairs of quadratures.

\section{Conclusions}

To conclude, we have studied the robust generation of photon entanglement with an electromechanical system, in which two CPWC's are capacitively coupled by a MR. With this cavity electromechanical system, we have presented two different schemes to generate two-mode continuous-variable entangled states of microwave photons confined in the cavities. The first scheme is based on coherent
control over the dynamics of the system to selectively induce excitations of the cavity Bogoliubov modes.
The second one is based on
a dissipative quantum dynamical process, which exploits the mechanical dissipation as a useful resource to implement ground state cooling of the cavity Bogoliubov modes. These protocols may have interesting applications in quantum information processing with
electromechanical systems.

\section*{Acknowledgments}
This work is supported by the Special Prophase Project in the
National Basic Research Program of China under Grant No.
2011CB311807, the NNSF of China under
Grant No. 11104215, and the Research Fund for the Doctoral
Program of Higher Education of China under Grant No.
20110201120035.

\emph{Note added}: After completing this work, we became aware of two
related works on the arXiv, which exploit the quite same ideas to generate photon entanglement
with optomechanical systems. These works have already been published \cite{prl-110-233602,prl-110-253601}.

%

\end{document}